# Photonic *p-n* Junction: An Ideal Near-Field Heat Flux Modulator

Deyu Xu[a], Junming Zhao[a,b,*], Linhua Liu[3]

[a]*School of Energy Science and Engineering, Harbin Institute of Technology,*
*Harbin 150001, China*

[b]*Key Laboratory of Aerospace Thermophysics, Ministry of Industry and Information Technology*,
*Harbin 150001, China*

[3] *School of Energy and Power Engineering, Shandong University,*
*Qingdao 266237, China*

**Abstract**

Using a pair of *p*- and *n*-type semiconductors separated by a nanoscale vacuum gap, we introduce an optoelectronics element prototype, "photonic *p-n* junction", as an analogue of the electronic *p-n* junction, which is demonstrated to serve as an ideal near-field heat flux modulator. The high modulation performance relies on the switch among three fundamental photon-carrier interaction modes (i.e., surface plasmon polaritons, symmetric depleted internal plasmon polaritons and symmetric accumulated surface plasmon polaritons) caused by the changes in hole and electron densities near the surfaces under a tunable bias. This prototype offers new thoughts not only for contactless thermal management at nanoscale but also for design of optoelectronics devices processing information carried by plasmon polaritons.

---

[*] Corresponding author.

*Email address*: jmzhao@hit.edu.cn

Modern electronics, which have fundamentally changed our lives, are thereby inspiring the emergences of thermal photonic analogues in the past decade, such as thermal diodes [1-3], transistors [4-7], memristor [8], logic gates [9,10] and other prototypes [11-16]. These inventions suggest the potential to process information via manipulating radiative transfer, which is of higher transport speed than thermal conduction and may be drastically enhanced utilizing evanescent photons in the near-field [17-20].

Among various functional manipulations based on near-field thermal photons, electrically modulating the heat flux is an important direction due to the simplicity and flexibility of applying bias voltage. Motivated by the electronic metal-oxide-semiconductor (MOS) transistor, Papadakis *et al*. [21] proposed a thermal MOS switch, which, under a bias, can tune the surface plasmon polaritons of the active region and in turn the near-field heat flux (then experimentally implemented by Thomas *et al*. [22] resorting to graphene as the semiconductor layer). The mechanism therein manifests that optoelectronics devices, especially those involving plasmonics [23-27], are promising candidates for high-speed heat flux modulation at nanoscale.

In this Letter, we are inspired by another fundamental element in electronics, the *p-n* junction, and propose a *photonic p-n junction*, which is composed of a pair of *p*- and *n*-type semiconductors separated by a nanoscale vacuum gap (see FIG. 1), to serve as a near-field heat flux modulator. We demonstrate that, thanks to the unique *p-n* cooperation configuration, a controllable bias can turn the surface plasmon polaritons into other two photon-carrier interaction modes (namely, symmetric depleted internal plasmon polaritons [28] or symmetric accumulated surface plasmon polaritons), therefore dramatically modulating the photon heat flux. We also show that it is not easy to obtain comparable performance using the *p-p* counterpart.

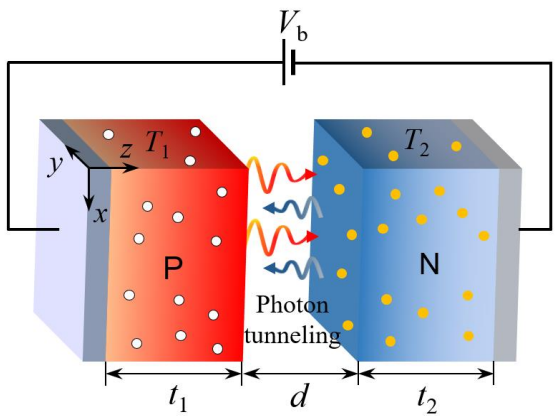

FIG. 1 Schematic of the photonic *p-n* junction consiting of a *p*-type semiconductor with thickness $t_1$ and a *n*-type semiconductor with thickness $t_2$, separated by a nanoscale vacuum gap of thickness $d$. A bias voltage $V_b$ is applied across the junction trough Ohmic contact, with the anode connected to the *p*-side. In this way, the photon heat flux $Q$ between the *p*-side maintained at temperature $T_1$ and the *n*-side maintained at temperature $T_2$ ($T_1 \geq T_2$) can be controlled as a function of $V_b$ ($V_b > 0$ for forward bias and $V_b < 0$ for reverse bias).

Let us begin with the concept of the photonic *p-n* junction. To give an embodiment showing how the heat flux through the photonic *p-n* junction can be modulated, we choose doped silicon (Si), the most common material in electronic industry, as both the *p*- and *n*-type semiconductors. Under a bias, holes in the *p*-Si and electrons in the *n*-Si are redistributed according to the Poisson equation: $\frac{d}{dz}\left(\varepsilon_{dc}\frac{d\psi}{dz}\right) = -e[N_D - N_A + p(z) - n(z)]$, where, $z$ is the location coordinate (see FIG. 1), $\psi$ is the $z$-dependent electric potential, $\varepsilon_{dc}$ is the DC permittivity, $e$ is unit electronic charge, $N_D$, $N_A$, $p$, $n$ are the densities of doping donors, doping acceptors, holes and electrons, respectively [29]. $p(z)$ and $n(z)$ can be linked to $\psi$ according to the Boltzmann statistics [29-31] and thus calculated by solving the Poisson equation via iteration methods [32]. The resultant distributions of dielectric function for both the *p*- and *n*-Si are described by the Drude model as [33,34]:

$$\varepsilon(\omega, z) = \varepsilon_\infty - \frac{\omega_{p,z}^2}{(\omega^2 + i\omega\Gamma_z)}, \qquad (1)$$

where, $\omega$ is the angular frequency, $\varepsilon(\omega, z)$ is the dielectric constant at frequency $\omega$ and location $z$, $\varepsilon_\infty$ is the high frequency dielectric constant, being 11.7 for Si. The plasma frequency $\omega_{p,z}$ is determined by the majority carrier density $N_{maj}(z)$ [= $p(z)$ for the *p*-Si while = $n(z)$ for the *n*-Si] through [25,35]

$$\omega_{p,z} = \sqrt{\frac{N_{maj}(z)e^2}{\varepsilon_0 m^*}}, \qquad (2)$$

where, $\varepsilon_0$ is the permittivity of free space and $m^*$ is the carrier effective mass. $\Gamma_z$ is the scattering rate which is also influenced by carrier density [33,34] and is therefore dependent on $z$. Metal electrodes made of aluminum serve as substrates whose dielectric functions are also described by the Drude model from Ref. [36].

When discussing heat transfer, both sides are approximated by a multilayer medium with the dielectric function of each layer determined by its $z$-location following Eqs. (1) and (2), which enables us to treat the involved gradient-property problem [28,32,37]. The local monochromatic heat flux in the *n*-side due to emitting of the *p*-side is given by the fluctuational electrodynamics with the help of dyadic Green's functions, as [28,38]

$$q_{\omega, p \to n}(z) = \frac{\omega^2 \Theta(\omega, T_1)}{\pi^2 c^2} \sum_s \text{Re}\left\{ i\varepsilon_s''(\omega, z_s) \int_0^\infty \beta d\beta \int_{z_s}^{z_{s+1}} dz' \begin{bmatrix} g_{sl\rho\rho}^E(\beta, z, z', \omega) g_{sl\theta\rho}^{H*}(\beta, z, z', \omega) \\ + g_{sl\rho z}^E(\beta, z, z', \omega) g_{sl\theta z}^{H*}(\beta, z, z', \omega) \\ - g_{sl\theta\theta}^E(\beta, z, z', \omega) g_{sl\rho\theta}^{H*}(\beta, z, z', \omega) \end{bmatrix} \right\}, \qquad (3)$$

where, $z$ is the interested location in the *n*-side. $\Theta(\omega, T) = \hbar\omega / [\exp(\hbar\omega / k_B T) - 1]$ is the mean energy of Planck oscillator at $\omega$ and $T$, $\hbar$ is the Planck constant divided by $2\pi$, $k_B$ is the Boltzmann constant. $\beta$ is the wavevector parallel to the interface. $z'$ represent the location of an emitting point in an arbitrary layer $s$ and the integration over $[z_s, z_{s+1}]$ gets the emission of the whole layer $s$. $\sum_s(\ )$ sums the contributions from all layers constituting the material. $g_{sl}^{E(H)}$ are the components of the Wely representations of electric (magnetic) dyadic Green's tensors, where $l$ is the layer that the interested point locates at. The formulas of $g_{sl}^{E(H)}$ can be found in Ref. [38]. The local monochromatic heat flux in the $p$-side due to emitting of the $n$-side $q_{\omega,n\to p}$ is determined symmetrically by Eq. (3).

Firstly, we inspect the local absorption distribution of spectral heat flux $|dq_\omega(z)/dz|$ to show the tunability of photon-carrier interaction in FIG. 2. To exclude the interfere of temperature difference when analyzing the effect of carriers on photon absorption, $T_1$ and $T_2$ are set equally to 300 K. At zero bias, the carriers in the $p$- and $n$-Si are distributed uniformly [see FIG. 2(a)], due to the flat energy band within both semiconductors ignoring the work function difference between them [29]. For the photonic $p$-$n$ junction, as shown in FIG. 2(b) and (c), holes in the $p$-Si and electrons in the $n$-Si synchronously deplete near the Si/vacuum interfaces (surfaces) at $V_b = -5$ V while synchronously accumulate there at $V_b = 20$ V. The changes in carrier densities are bound to modulate the photon-carrier interactions.

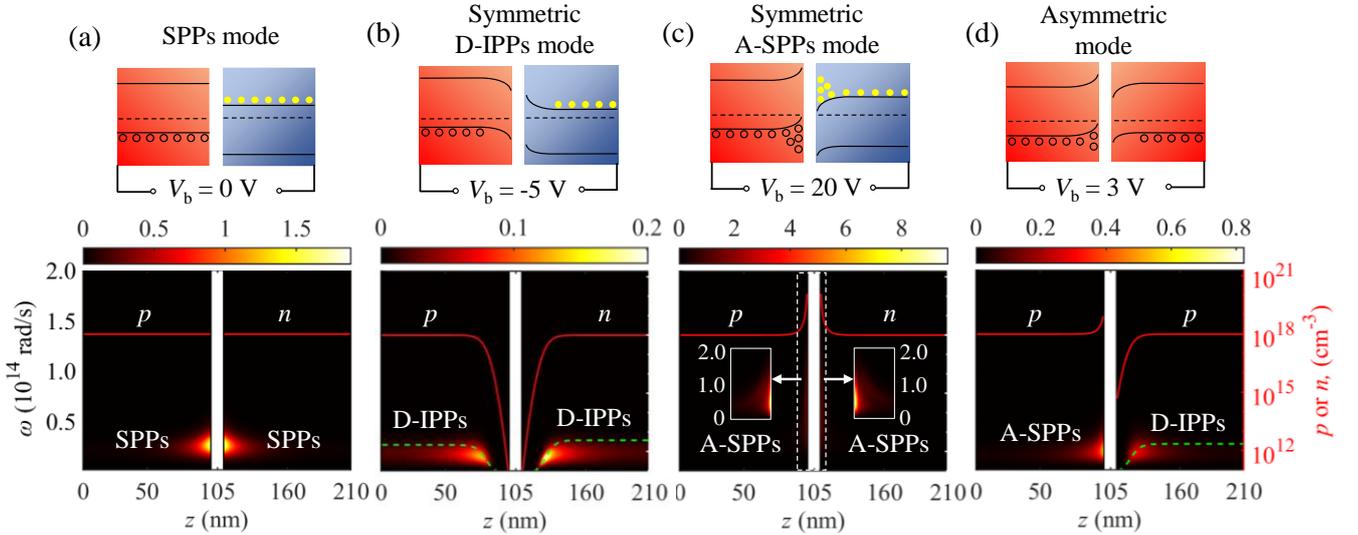

FIG. 2 Local absorption distributions of spectral heat flux $|dq_\omega/dz|$ (Wm$^{-3}$rad$^{-1}$s) in both sides of the vacuum gap for the photonic $p$-$n$ junctions at (a) zero bias, (b) $V_b = -5$ V, (c) $V_b = 20$ V and for (d) the photonic $p$-$p$ junction at $V_b = 3$ V. $T_1 = T_2 = 300$ K. The equilibrium majority carrier densities are set equally to $N_b = 10^{18}$ cm$^{-3}$ for both sides and other parameters are set to $t_1 = t_2 = 100$ nm, $d = 10$ nm. These parameters are used throughout this Letter unless otherwise specified. In the right vertical axes, the corresponding majority carrier densities are plotted with red lines. The green dash lines in (b) and (d) present the $\varepsilon = 0$ isolines given by Eq. (4). The upper insets sketch the energy band diagrams for corresponding cases that cause carrier accumulation or depletion (the dash lines denote the Fermi levels).

At zero bias, the thermal photons are absorbed strongly near the surfaces due to the excitation of surface plasmon polaritons (SPPs) (further verified by FIG. 4). At $V_b = -5$ V, however, the absorption near the surfaces are severely suppressed, and instead, the absorption in the interior is enhanced. It can be clearly seen that the locations of the strongest absorption for the photons with frequency $\omega$ agree well with the implicit $z$-$\omega$ relation of

$$N_{\text{maj}}(z) = \frac{\varepsilon_0 \varepsilon_\infty m^* \omega^2}{e^2} \tag{4}$$

shown with the green dash lines in FIG. 2(b), which is obtained by zeroing Eq. (1). This illustrates that internal plasmon polaritons (IPPs) dominate the photon absorption in this case [28]. The gradient distribution of dielectric function caused by the gradient depletion of carriers provides internal "interfaces" for plasmon polaritons. In terms of space distribution, the plasmon polaritons "deplete" near the surfaces following the depletion of carriers and are shifted into the interior of materials, symmetrically for both sides of the vacuum gap. We therefore name this photon-carrier interaction of the system as symmetric *depleted internal plasmon polaritons* (D-IPPs) mode.

The scenario of $V_b$ = 20 V is completely different, as shown in FIG. 2(c) and the enlarged views of near-surface regions. It is intuitively observed that, for both sides of the vacuum gap, the heat flux is almost fully absorbed in the proximity of surface, as if a two-dimensional plasma layer (e.g., graphene) are inserted which largely confines the radiative energy to the surface [39,40] [compared with FIG. 2(a)]. As this photon-carrier interaction mode is caused by the accumulation of carriers, we name it as symmetric *accumulated surface plasmon polaritons* (A-SPPs) mode. In this mode, more photons of higher frequencies are effectively absorbed, for which the mechanism will be uncovered later. For the case of photonic *p-p* junction at $V_b$ = 3 V shown in FIG. 2(d), holes accumulate at the surface for the *p*-Si but deplete for the *n*-Si, causing the asymmetric mode which supports A-SPPs and D-IPPs in the two sides, respectively.

Above, we have recognized four kinds of photon-carrier interaction modes. Their effects on near-field heat transfer are ready to be studied. We exhibit in FIG. 3 the exchanged heat flux through the photonic *p-n* and *p-p* junctions under a temperature difference of $T_1$ = 310 K, $T_2$ = 290 K, as a function of bias voltage in the range [-20, 20] V. This voltage range is chosen as an example here and depends on the breakdown electric field in practical applications. Let us firstly focus on the region covered by the yellow shadow in FIG. 3. For the *p-n* junction, symmetric D-IPPs gradually convert to SPPs as $V_b$ increases from -6.5 V to 0 and then to symmetric A-SPPs as $V_b > 0$. The heat transfer is resultantly enhanced. Higher positive $V_b$ enhances A-SPPs and further promotes the heat flux. For the *p-p* junction, in contrast, either the positive or negative $V_b$ reduces the heat flux, which can be explained as follows taking the case of $V_b$ = 3 V as an example (obviously, given the inversion invariance of the *p-p* junction, opposite $V_b$ play same roles as demonstrated by the symmetry of its $Q$-$V_b$ curve). Although A-SPPs can be excited in one side, these electromagnetic states cannot be fully absorbed by the other side which only supports D-IPPs [see FIG.

2(d)]. In this voltage region, the photonic *p-n* junction exhibits significant superiority in heat flux modulation with a modulation ratio [$\Phi = (Q_{max} - Q_{min})/Q_0$, where $Q_0$ is the heat flux at zero bias] of 1.49, much higher than that of the *p-p* junction, 0.84.

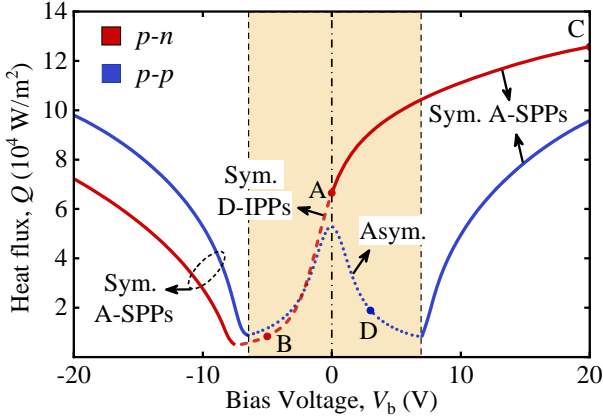

FIG. 3 Exchanged heat flux $Q$ as a function of bias voltage $V_b$ for the photonic *p-n* and *p-p* junctions in the range [-20, 20] V. $T_1$ = 310 K and $T_2$ = 290 K. Different line styles rerepresent the regions dominated by different photon-carrier modes. The four cases in FIG. 2 are marked with A, B, C, D, respectively. The shadow area denotes the vloatge region within which $Q$ of the *p-p* junction decreases with $|V_b|$ and the *p-n* juncion exhibits significant superiority in heat flux modulation.

At reverse bias and $|V_b|$ is higher than a critical value such that $|\psi_s| > 2\ln(N_b/n_i)$ ($\psi_s$ is the dimensionless surface potential, $n_i$ is the intrinsic carrier density) for at least one side of the photonic *p-n* junction, the density of minority carrier at the surface would exceed $N_b$ [30]. Then, the A-SPPs are again excited by the interactions of photons and accumulated minority carriers. As a result, $Q$ increases with $|V_b|$ like the case of forward bias. Similar scenarios occur for the *p-p* junction where asymmetric mode switch to symmetric A-IPPs mode at high $|V_b|$ as shown by blue solid lines in FIG. 3. Although relatively large modulation range can be obtained by using the *p-p* junction in this region, it needs higher $|V_b|$ which increases the risk of breakdown.

Up to now, one remaining issue is the electromagnetic state natures in the SPPs, symmetric D-IPPs mode, symmetric A-SPPs mode and asymmetric mode, which are responsible for their effects on heat transfer as shown above. To this end, we investigate the *p*-polarized photon tunneling probability for evanescent waves ($\beta > k_0 = \omega/c$) [41,42], which is written as

$$\xi_p(\omega,\beta) = \frac{4\,\text{Im}(R_{01}^p)\,\text{Im}(R_{02}^p)e^{-2\text{Im}(\gamma_0)d}}{\left|1 - R_{01}^p R_{02}^p e^{2i\gamma_0 d}\right|^2}, \qquad (5)$$

where $\gamma_0 = \sqrt{k_0^2 - \beta^2}$ is the wavevector perpendicular to the interface in vacuum. $R_{0i}^p$ is the reflection coefficient for *p*-polarization between vacuum and medium *i*, which can be obtained by transfer matrix method from the dielectric function distribution [32,38,43].

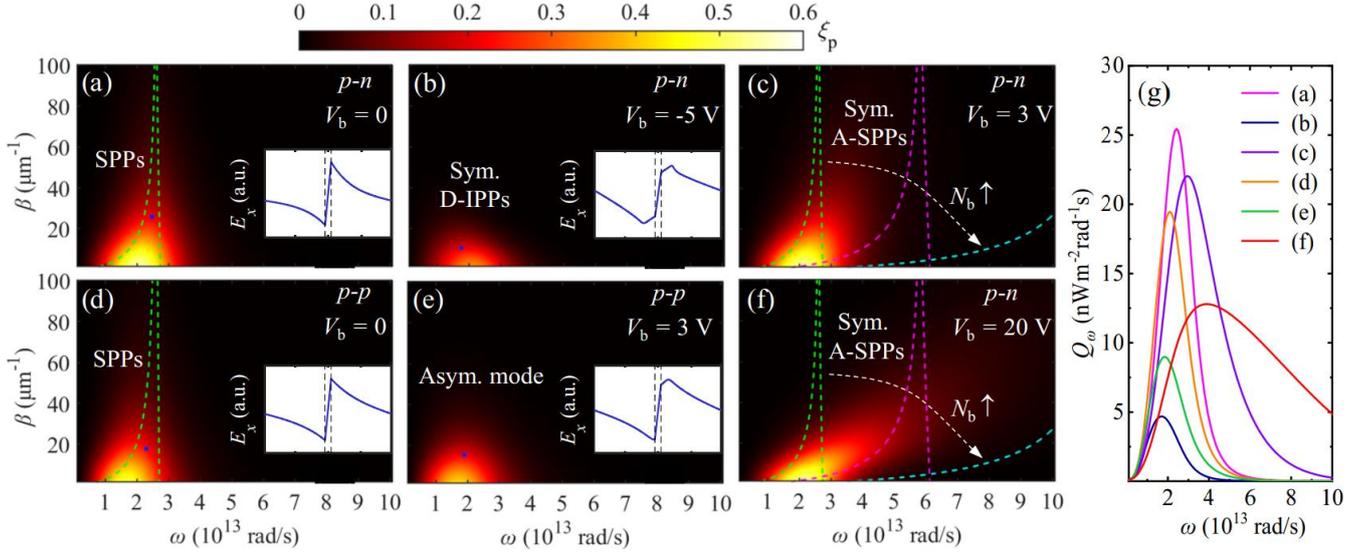

FIG. 4 Electromagnetic states in various modes that contribute to heat trasfer. (a)-(f) *p*-polarized photon tunneling probabilities ($\xi_p$) in the $\omega$ - $\beta$ planes for corresponding cases. The green dash lines, pink dash lines and cyan dash lines represent the dispersion relations of SPPs for $N_b = 10^{18}$ cm$^{-3}$, $5\times10^{18}$ cm$^{-3}$ and $2\times10^{19}$ cm$^{-3}$, determined by Eq. (7). The white dash arrows illustrate the evolution trend of SPPs as $N_b$ increases. In the insets of panels (a), (b), (d), (e), the tangential electric fields $E_x$ in the corresponding system for the typical electromagnetic states marked with blue dots are plotted and the dash lines denote the material/vacuum interfaces. (g) Exchanged spectral heat fluxes $Q_\omega$ for the cases in (a)-(f).

FIG. 4(a)-(f) show $\xi_p$ in the $\omega$ - $\beta$ plane for six typical cases. For the *p*-Si / vacuum / *n*-Si (*p*-Si) sandwich system, the dispersion relation of SPPs can be derived by applying boundary continuity conditions on the evanescent electromagnetic fields [35,44], which yields

$$\frac{\gamma_0\varepsilon_1-\gamma_1}{\gamma_0\varepsilon_1+\gamma_1}\frac{\gamma_0\varepsilon_2-\gamma_2}{\gamma_0\varepsilon_2+\gamma_2}=e^{2\mathrm{Im}(\gamma_0)d}, \quad (6)$$

where $\varepsilon_1$ and $\varepsilon_2$ are respectively the dielectric functions of the two sides of the vacuum gap, $\gamma_i=\sqrt{\varepsilon_i k_0^2-\beta^2}$ ($i = 1, 2$). For $\beta \gg \omega/c$, $\gamma_i \approx i\beta$, and the wavevector of SPPs can hence be written as the explicit function of $\omega$ as

$$\beta_{\mathrm{SPP}}=\frac{1}{2d}\ln\left(\frac{\varepsilon_1-1}{\varepsilon_1+1}\frac{\varepsilon_2-1}{\varepsilon_2+1}\right), \quad (7)$$

derived from Eq. (6). The dispersion relations determined by Eq. (7) are plotted with green dash lines in FIG. 4(a) and (d), respectively for the cases of *p-n* junction and *p-p* junction at zero bias. They agree with the area of nonzero $\xi_p$ in which such electromagnetic states can tunnel the vacuum gap and contribute to heat transfer.

Shown in FIG. 4(b) is the electromagnetic states in the symmetric D-IPPs mode, which are of lower frequencies and smaller wavevectors compared to the SPPs. The electric fields take their peaks at the interior of both sides for symmetric D-IPPs mode [inset in FIG. 4(b)], rather than at the surfaces as is the case in SPPs shown in FIG. 4(a) and (d). This is the natural characteristic of IPPs which has been introduced in Ref. [28], and here, we find they can be excited in a realistic structure. The red-shifted frequencies can be understood by considering the condition of IPPs [green dash lines in FIG. 2(b)], and the reduced wavevectors are explained by the longer distance of heat transfer caused by the non-absorbing near-surface regions [see FIG. 2(b)].

The symmetric A-SPPs extend their frequency band to higher region with sacrificing the magnitudes of wavevectors compared to SPPs, as shown in FIG. 4(c), which is due to the coupling of SPPs supported by the regions with accumulated carriers. A higher carrier density "pulls" the SPPs to the higher frequencies, due to the increase in $\omega_p$ [according to Eq. (2)], while $\beta$ is reduced for a given $\omega$ [illustrated with the white dash arrows in FIG. 4(c) and (f)]. This agrees with the evolution trend of the nonzero-$\xi_p$ area as $V_b$ varies from 3 V to 20 V [compare FIG. 4(c) and (f)], along with which the carriers near the surfaces are increasingly accumulated. FIG. 4(e) shows the asymmetric mode for the *p-p* junction, whose states are limited by the D-IPPs, thus suppressing the heat transfer compared with the zero-bias case [FIG. 4(d)]. The electric field distribution visually illustrates the asymmetry. Above analysis explains the variations of the exchanged spectral heat flux peaks as shown in FIG. 4(g).

In conclusion, we proposed a new optoelectronics element prototype, "photonic *p-n* junction", ideal for electrically modulating near-field heat transfer. Photon-carrier interactions are driven to interconvert among SPPs, symmetric D-IPPs mode that suppresses photon tunneling and symmetric A-SPPs mode that promotes photon tunneling, by the change in carrier densities near the surfaces under bias voltage, yielding a considerable modulation in heat flux. The photonic *p-p* junction cannot perform so well as it can only support asymmetric mode which suppresses heat transfer at moderate voltages. The conclusion can be extended to the *n-n* junction. Although this Letter focuses on the modulation of heat flux, the proposed mechanism provides new insights for future plasmonic tuning optoelectronics.

The support of this work by the National Natural Science Foundation of China (No. 51976045) is gratefully acknowledged. JMZ also thanks Prof. Philippe Ben-Abdallah for very valuable discussions of the work.